\documentclass[journal, twocolumns]{IEEEtran}

\usepackage{epsfig}
\usepackage{graphicx}
\usepackage{dcolumn}
\usepackage{bm}
\usepackage{epsfig}
\usepackage{epsfig}
\usepackage{amsmath}
\usepackage{amsfonts}

\newcommand{\ba}{\begin{eqnarray}}
\newcommand{\ea}{\end{eqnarray}}
\newcommand{\beq}{\begin{equation}}
\newcommand{\eeq}{\end{equation}}
\newcommand{\bea}{\begin{eqnarray}}
\newcommand{\eea}{\end{eqnarray}}

\usepackage{theorem}

\newtheorem{definition}{Definition}

\newtheorem{proposition}{Proposition}

\theorembodyfont{\rmfamily}
\newtheorem{remark}{Remark}
\theoremstyle{break}

\def\QED{~\rule[-1pt]{5pt}{5pt}\par\medskip}

\def\n{\noindent}

\def\Re {\mathbb{R}}

\begin{document}
%
\title{Characterization of majorization monotone quantum dynamics}
%
%
\author{Haidong Yuan 
\thanks{Haidong Yuan is with the department of Mechanical Engineering, Massachusetts Institute of Technology.}
}
\maketitle

\begin{abstract}
In this article I study the dynamics of open quantum system in
Markovian environment. I give necessary and sufficient conditions for such dynamics to be
majorization monotone, which are those dynamics always mixing the states.
\end{abstract}



%
\IEEEpeerreviewmaketitle

\section{\label{sec:introduction}Introduction}
In the last two decades, control theory has been applied to an
increasingly wide number of problems in physics and chemistry
whose dynamics are governed by the time-dependent Schr\"odinger
equation (TDSE), including control of chemical reactions
\cite{Tannor85,Tannor86,Tannor88,Kosloff89,Rice2000,Shapiro03,Brixner03,Mitric02},
state-to-state population transfer
\cite{Peirce88,Shi91,Tannor04,Jakubetz90,Shen94}, shaped wavepackets
\cite{Yan93}, NMR spin dynamics \cite{Khaneja01, Yuan00,Yuan01,Yuan02,Yuan03}, Bose-Einstein
condensation \cite{Hornung,Sklarz02.1,sklarz02.2}, quantum
computing \cite{Rangan01,Tesch01,Palao02,Yuan04,Yuan05}, oriented rotational
wavepackets \cite{Leibscher03,Rabitz,Gordon97}, etc. More
recently, there has been vigorous effort in studying the control
of open quantum systems which are governed by Lindblad equations,
where the central object is the density matrix, rather than the
wave function
\cite{Bartana93,Bartana,TannorRot,Cao97,Gross98,Ohtsuki03,Khaneja03}.
The Lindblad equation is an extension of the TDSE that allows for
the inclusion of dissipative processes. In this article, I will study
those dynamics governed by Lindblad equations and give necessary and
sufficient conditions for the dynamics to be majorization monotone, which are those
dynamics always mixing the states. This study suggests that majorization may serve as time arrow under these dynamics in analog to entropy in second law of thermal dynamics.

The article is organized as following: section \ref{sec:majorization} gives a brief introduction to majorization;
section \ref{sec:quantum} gives the definition of majorization monotone quantum dynamics; then in section \ref{sec:result},
necessary and sufficient conditions for majorization monotone quantum dynamics are given.

\section{\label{sec:majorization}Brief introduction to majorization}
In this section I give a brief introduction on majorization, most stuff in this section can be found in the second chapter of Bhatia's book \cite{Bhatia97}.

For a vector $x=(x_1 , . . . ,x_n)^T$ in
$\Re^n$, we denote by $x^\downarrow =(x_1^\downarrow , . . .
,x_n^\downarrow)^T$ a permutation of $x$ so that
$x_i^\downarrow\geq x_j^\downarrow$ if $i<j$, where $1\leq i,
j\leq n$.
\begin{definition}[majorization]{\rm
A vector $x\in \Re^n$ is majorized by a vector $y\in \Re^n$
(denoted by $x \prec y$), if
\begin{equation}
  \sum_{j=1}^d x^{\downarrow}_j \leq \sum_{j=1}^d y^{\downarrow}_j
\end{equation}
for $d = 1,\ldots,n-1$, and the inequality holds with equality
when $d= n$.}
\end{definition}

\begin{proposition}
\label{prop:convexhull} {\rm $x \prec y$ iff $x$ lies in the
convex hull of all $P_iy$, where $P_i$
are permutation matrices.}
\end{proposition}

\begin{proposition}
\label{prop:double} {\rm $x \prec y$ if and only if $x=Dy$ where D is doubly stochastic matrix.}
\end{proposition}
\begin{remark}{\rm A doubly stochastic matrix D is a matrix with non-negative entries and every column and row sum to 1, i.e., $d_{ij}\geq 0$, $\sum_i d_{ij}=1$, $\sum_j d_{ij}=1$.}\end{remark}

\begin{proposition}
\label{prop:cov}{\rm Suppose $f$ is a convex function on $\Re$, and $x\prec y$ in $\Re^n$, then $$\sum_i^n f(x_i)\leq \sum_i^n f(y_i).$$}
\end{proposition}

\begin{proposition} \label{prop:Schur} {\rm For a vector
$\lambda=(\lambda_1 , . . . ,\lambda_n)^T$, denote $D_\lambda$ a
diagonal matrix with $(\lambda_1 , . . . ,\lambda_n)$ as its
diagonal entries, let $a=(a_1,...,a_n)^T$ be the diagonal entries
of matrix $A=K^TD_\lambda K$, where $K\in SO(n)$. Then $a \prec
\lambda$. Conversely for any vector $a \prec \lambda$, there
exists a $K\in SO(n)$, such that $(a_1,...,a_n)^T$  are the
diagonal entries of $A=K^TD_\lambda K$.}
\end{proposition}

\begin{remark}{\rm $SO(n)$ is the group of special orthogonal matrices, $K\in SO(n)$ means $K^TK=I$ and $det(K)=1$.}\end{remark}

\section{Majorization in open quantum dynamics}
\label{sec:quantum}
The state of an open quantum system of N-level can be represented by a $N\times N$ positive semi-definite, trace $1$ matrix, called density matrix.
Let $\rho$ denote the density matrix of an quantum system, its dynamics in markovian environment is governed by the Lindblad equation, which takes
the form
\begin{equation}\label{eq:main}
\dot{\rho} = -i[H(t), \rho] + L(\rho),
\end{equation}where $-i[H, \rho]$ is the unitary evolution of the
quantum system and $L(\rho)$ is the dissipative part of the
evolution. The term $L(\rho)$ is linear in $\rho$ and is given by
the Lindblad form \cite{Lindblad,Alicki86},
$$L(\rho) = \sum_{\alpha \beta}a_{\alpha \beta}(F_{\alpha}\rho F_{\beta}^{\dagger} - \frac{1}{2}
\{F_{\beta}^{\dagger}F_{\alpha}, \rho \}) , $$ where $F_{\alpha},
F_{\beta}$ are the Lindblad operators, which form a basis of $N\times N$ trace $0$ matrices (we have $N^2-1$ of them) and $\{A,B\}=AB+BA$. If we put the coefficient $a_{\alpha\beta}$ into a $(N^2-1)\times (N^2-1)$ matrix $A=(a_{\alpha\beta})$, it is
known as the GKS (Gorini, Kossakowski and Sudarshan)
matrix~\cite{GKW}, which needs to be positive semi-definite.

Eq. (\ref{eq:main}) has
the following three well known properties: 1) ${\rm Tr}(\rho)$
remains unity for all time, 2) $\rho$ remains a Hermitian matrix,
and 3) $\rho$ stays positive semi-definite, i.e. that
$\rho$ never develops non-negative eigenvalues.

\begin{definition}{\rm Suppose $\rho_1$ and $\rho_2$ are two states of a quantum system, we say $\rho_1$ is majorized by $\rho_2$ ($\rho_1 \prec \rho_2$) if the eigenvalues of $\rho_1$ is majorized by the eigenvalues of $\rho_2$ ($\lambda(\rho_1)\prec \lambda(\rho_2)$).}
\end{definition}

Basically majorization gives an order of mixed-ness of quantum states, i.e., if $\rho_1\prec \rho_2$, then $\rho_1$ is more mixed than $\rho_2$, which can be seen from the following propositions.

\begin{definition}[Von Neumann entropy]{\rm The Von Neumann entropy of a density matrix is given by $$S(\rho)=-Tr[\rho log(\rho)].$$}
\end{definition}

\begin{proposition}\label{prop:entropy}{\rm If $\rho_1\prec \rho_2$, then $S(\rho_1)\geq S(\rho_2)$.}
\end{proposition}

\begin{proposition}\label{prop:trace}{\rm If $\rho_1\prec \rho_2$, then $Tr(\rho_1^2)\leq Tr(\rho_2^2)$.}
\end{proposition}
\begin{remark}{\rm The above two propositions can be easily derived from Proposition ~\ref{prop:cov}.}\end{remark}
The entropy and trace norm are usually used to quantify how mixed quantum states are. But majorization is a more strong condition than these two functions, and in some sense it gives a more proper order of mixed-ness as we can see from the following proposition.

\begin{proposition}~\cite{Uhlmann} {\rm  $\rho_1 \prec \rho_2$ if and only if $\rho_1$ can be obtained by mixing the unitary conjugations of $\rho_2$, i.e., $\rho_1=\sum_i p_iU_i\rho_2 U_i^\dagger$, where $p_i>0, \sum_i p_i=1$ and $U_i$ are unitary operators.}
\end{proposition}

%
\section{Necessary and sufficient condition of majorization monotone quantum dynamics}
\label{sec:result}
\begin{definition}[Majorization monotone dynamics]{\rm An open quantum dynamics governed by Eq. (\ref{eq:main}) is majorization monotone if and only if $\rho(t_2)\prec \rho(t_1)$ when $t_2>t_1$, $\forall t_1, t_2$.}
\end{definition}
Intuitively majorization monotone dynamics are those kind of dynamics which always mixing the states. As we can see from Proposition ~\ref{prop:entropy}, these kind of dynamics always increase the entropy of the system. One can immediately see a necessary condition for a dynamics to be majorization monotone: the state $\rho_I=\frac{1}{N}I$ has to be a steady state of such dynamics, where $I$ is identity matrix. As $\frac{1}{N}I$ is the most mixed state, any state $\rho_i \prec \frac{1}{N}I$ would imply $\rho_i=\frac{1}{N}I.$ The question now is whether this condition is also sufficient.

Let's first look at a simple system: a single spin in markovian environment.

\subsection{An example on single spin}
\label{sec:example}
Take the general expression of the master equation
\begin{equation}
\label{eq:master}
\dot{\rho}=-i[H,\rho]+L(\rho),
\end{equation}

\n where
$$L(\rho)=\sum_{\alpha\beta}a_{\alpha\beta}(F_{\alpha}\rho
F_{\beta}^{\dagger}-\frac{1}{2}\{F_{\beta}^{\dagger}F_{\alpha},\rho\}).$$
\n For the single spin, we can take the basis $\{F_{\alpha}\}$ as
normalized Pauli spin operators
$\frac{1}{\sqrt{2}}\{\sigma_x,\sigma_y,\sigma_z\}$, where $\sigma_x=\left(\begin{smallmatrix}
0 & 1 \\
1 & 0
\end{smallmatrix}
\right)$, $ \sigma_{y}= \left(
\begin{smallmatrix}
0 & -i \\
i & 0
\end{smallmatrix}
\right) $, and $ \sigma_{z}= \left(
\begin{smallmatrix}
1 & 0 \\
0 & -1
\end{smallmatrix}
\right) $. The
coefficient matrix $$A=\left(\begin{array}{ccc}
       a_{xx} & a_{xy} & a_{xz} \\
       a_{yx} & a_{yy} & a_{yz} \\
       a_{zx} & a_{zy} & a_{zz}
\end{array}\right)$$
      is positive semi-definite.

 If identity state is a steady state, the right hand side of Eq. (\ref{eq:master}) should be $0$ when $\rho=\frac{1}{N}I$. As $-i[H,\frac{1}{N}I]=0$, so the condition reduces to $L(I)=0$ (since $L(\rho)$ is a linear map, we can ignore the constant $\frac{1}{N}$). This is equivalent to
\begin{equation}\label{eq:L}
\sum_{\alpha,\beta}a_{\alpha\beta}[F_\alpha,F_\beta^\dagger]=0.
\end{equation}
In the single spin case, substitute $F$ by Pauli matrices and it is easy to see that the above condition reduces to $$a_{\alpha\beta}=a_{\beta\alpha},$$ i.e., the GKS matrix should be real symmetric, positive semi-definite matrix \cite{bacon}, while the general GKS matrix is Hermitian, positive semi-definite. We want to see whether the dynamics of single spin under this condition is majorization monotone.

As majorization monotone is defined by the eigenvalues of density matrices, we are going to focus on the dynamics of the eigenvalues of density matrix.
Let $\Lambda$ be its associated
diagonal form of eigenvalues of density matrix $\rho$, i.e., $\Lambda$ is a diagonal matrix with eigenvalues of $\rho$ as its diagonal entries. At each instant of time, we can diagonalize the density matrix $\rho(t)=U(t)\Lambda(t)U^\dag(t)$ by a unitary matrix $U(t)$.

Substitute $\rho(t)=U(t)\Lambda(t)U^\dag(t)$ into
Eq. (\ref{eq:main}), we get \bea \aligned
\dot{\rho}(t)=&\dot{U}(t)\Lambda(t)U^\dag(t)+U(t)\dot{\Lambda}(t)U^\dag(t)\\&+U(t)\Lambda(t)\dot{U}^\dag(t)
\\=&-iH'(t)U(t)\Lambda(t)U^\dag(t)+U(t)\dot{\Lambda}(t)U^\dag(t)\\&+U(t)\Lambda(t)U^\dag(t)iH'(t)
\\=&-i[H'(t),U(t)\Lambda(t)U^\dag(t)]+U(t)\dot{\Lambda}(t)U^\dag(t)
\\=&-i[H(t),U(t)\Lambda(t)U^\dag(t)]+L[U(t)\Lambda(t)U^\dag(t)],
\endaligned
\eea where $H'(t)$ is defined by $\dot{U}(t)=-iH'(t)U(t)$, which is Hermitian. We
obtain \bea \aligned
\dot{\Lambda}(t)=&U^\dag(t)\{-i[H(t)-H'(t),U(t)\Lambda(t)U^\dag(t)]\\&+L[U(t)\Lambda(t)U^\dag(t)]\}U(t)
\\=&-i[U^\dag(t)(H(t)-H'(t))U(t),\Lambda(t)]\\&+U^\dag(t)L[U(t)\Lambda(t)U^\dag(t)]U(t).
\endaligned
\eea Note that the left side of the above equation is a diagonal
matrix, so for the right side we only need to keep the diagonal
part. It is easy to see that the diagonal part is zero for the
first term, thus we get \bea
\dot{\Lambda}(t)=diag(U^\dag(t)L[U(t)\Lambda(t)U^\dag(t)]U(t)), \eea
where we use $diag(M)$ to denote a diagonal matrix whose diagonal
entries are the same as matrix $M$.

%
%
%
%


\bea \label{eq:dyna1}\aligned \dot{\Lambda}(t)=diag(U^\dag&
L(U\Lambda U^\dag)U)
             \\=diag(\sum_{\alpha \beta}&a_{\alpha\beta}(U^\dag
             F_{\alpha}U\Lambda U^\dag F_{\beta}^\dag U\\&-\frac{1}{2}\{U^\dag
             F_{\beta}^\dag UU^\dag
             F_{\alpha}U, \Lambda\}))
             \\=diag(\sum_{\alpha \beta}&a_{\alpha\beta}(U^\dag
             F_{\alpha}U\Lambda U^\dag F_{\beta} U\\&-\frac{1}{2}\{U^\dag
             F_{\beta} UU^\dag
             F_{\alpha}U, \Lambda\})).
\endaligned
\eea For the last step we just used the fact that $F_{\beta}$ is a
Pauli matrix which is Hermitian. Now
$$U^\dag F_{\alpha}U=c_{\alpha \gamma}F_{\gamma},$$
where $C=\left(\begin{array}{ccc}
       c_{xx} & c_{xy} & c_{xz} \\
       c_{yx} & c_{yy} & c_{yz} \\
       c_{zx} & c_{zy} & c_{zz}

      \end{array}\right)\in SO(3)$ is the adjoint representation
      of $U$. Substituting these expressions into Eq. (\ref{eq:dyna1}), we
      obtain

\bea \label{eq:dyna2}\aligned \dot{\Lambda}(t)
             =diag(\sum_{\alpha \beta}a'_{\alpha\beta}(
             F_{\alpha}\Lambda F_{\beta} -\frac{1}{2}\{
             F_{\beta}F_{\alpha}, \Lambda\})),
\endaligned
\eea where $$a'_{\alpha \beta}=c_{\gamma \alpha}a_{\gamma
\mu}c_{\mu \beta}$$ are entries of transformed GKS matrix,
$$A'=C^TAC.$$
We can write $\Lambda(t)=\left(\begin{smallmatrix}
\frac{1}{2}+\lambda(t) & 0 \\
0 & \frac{1}{2}-\lambda(t)
\end{smallmatrix}
\right)=\frac{1}{2}I+\lambda(t)\sigma_z$, where $\lambda(t)\in [0,\frac{1}{2}]$. Substitute it into
Eq. (\ref{eq:dyna2}), we obtain the dynamics for
$\lambda(t)$,
$$\dot{\lambda}(t)=-(a'_{11}+a'_{22})\lambda(t).$$
From Proposition ~\ref{prop:Schur}, we
know
$$\mu_3+\mu_2\leq a'_{11}+a'_{22}\leq \mu_2+\mu_1,$$ where
$\mu_1\geq \mu_2\geq\mu_3$ are eigenvalues of the GKS matrix. From
this it is easy to see that at time $T$, the value of
$\lambda(T)$ lies in the following interval
$$[e^{-(\mu_1+\mu_2)T}\lambda(0),e^{-(\mu_2+\mu_3)T}\lambda(0)],$$
which is always less or equal to $\lambda(0)$. And this is sufficient for the dynamics to be majorization monotone in the single spin case, as $(\frac{1}{2}+\lambda(T),\frac{1}{2}-\lambda(T))\prec (\frac{1}{2}+\lambda(0),\frac{1}{2}-\lambda(0))$ when $\lambda(T)\leq \lambda(0)$. So $L(I)=0$ is necessary and sufficient for the dynamics to be majorization monotone in the single spin case. And this condition holds even we have coherent control on the spin, as the identity state remains as steady state in the presence of control as we can see from the following controlled dynamics:
$$\rho=-i[H(t)+\sum_i u_iH_i,\rho]+L(\rho),$$
where $\sum_iu_iH_i$ are our coherent controls. Suppose the controllers are able to generate any unitary operations on the spin fast compare to the dissipative rate, then $\lambda(T)$ can actually take any value in the interval $[e^{-(\mu_1+\mu_2)T}\lambda(0),e^{-(\mu_2+\mu_3)T}\lambda(0)]$, i.e., the reachable set for the single spin under the controlled Lindblad dynamics is \bea \aligned \nonumber
\rho(T)&=\{U\left(\begin{array}{cc}
       \frac{1}{2}+\lambda(T) & 0 \\
       0 & \frac{1}{2}-\lambda(T)  \\
         \end{array}\right)U^\dag |
\lambda(T)\in
\\&[e^{-(\mu_1+\mu_2) T}\lambda(0),e^{-(\mu_2+\mu_3)
T}\lambda(0)], U\in SU(2)\}.
\endaligned
\eea

This is to say that although controls can't reverse the direction of mixing, it can change the rate within some region.

\begin{remark}{\rm From the single spin case, we can see that a dynamical system being majorization monotone does not imply the states of the system always converge to identity state, as identity being a steady state does not exclude the possible existence of other steady states, for example, the dynamics given by $$\dot{\rho}=-i[\sigma_z,\rho]+\gamma[\sigma_z,[\sigma_z,\rho]],$$which describes the transverse relaxation mechanism in NMR, satisfies the majorization monotone condition in the single spin case, and it is easy to see that the state of this system does not necessary converge to identity matrix, in fact it can well be converged to any state of the form $\frac{1}{2}I+\alpha \sigma_z$, where $\alpha \in [0,\frac{1}{2}]$.}\end{remark}

\subsection{General Case}
In this section, we will show that $L(I)=0$ is also sufficient for the dynamics to be majorization monotone in the general case.
Suppose we solved the Lindblad equation
\begin{equation}\label{eq:main2}
\dot{\rho} = -i[H(t), \rho] + L(\rho),
\end{equation}
integrated this equation from $t_1$ to $t_2$, where $t_2>t_1$, and get a map: $$\Psi:\rho(t_1)\rightarrow \rho(t_2).$$ Such a map has a Kraus operator sum representation~\cite{Alicki86,Kraus83}:
\begin{equation}
\label{eq:Kraus}
\rho(t_2)=\Psi(\rho(t_1))=\sum_iK_i\rho(t_1)K_i^\dagger,
\end{equation}
 where in our case $\{K_i\}$ are $N\times N$ matrices, which depend on the dynamical Eq. (\ref{eq:main2}) and the time difference between $t_1$ and $t_2$. Also the Kraus operator sum has to be trace preserving as the trace of density matrix is always $1$, which implies that $$\sum_iK_i^\dag K_i=I.$$
If we have additional condition that identity state is a steady state of this dynamics, which means if $\rho(t_1)=\frac{1}{N}I$ then $\rho(t_2)$ remains at $\frac{1}{N}I$, substitute them into the Kraus operator sum representation, we will get an extra condition $$\sum_i K_iK_i^\dagger=I.$$
We will show these two conditions are enough to ensure the dynamics to be majorization monotone.
First let's diagonalize $\rho(t_1)$ and $\rho(t_2)$:
$$\rho(t_1)=U_1\Lambda(\rho(t_1))U_1^\dagger,$$
$$\rho(t_2)=U_2\Lambda(\rho(t_2))U_2^\dagger,$$
where $\Lambda(\rho)$ are diagonal matrix with eigenvalues of $\rho$ as its diagonal entries, substitute them into Eq. (\ref{eq:Kraus}), we get
\bea \aligned
U_2\Lambda(\rho(t_2))U_2^\dagger&=\sum_iK_iU_1\Lambda(\rho(t_1))U_1^\dagger K_i^\dagger,\\
\Lambda(\rho(t_2))&=\sum_i U_2^\dagger K_iU_1\Lambda(\rho(t_1))U_1^\dagger K_i^\dagger U_2.\\
  \endaligned
  \eea
  Let $V_i=U_2^\dagger K_iU_1$, then
  \begin{equation}
  \label{eq:Kraus-eigen}
  \Lambda(\rho(t_2))=\sum_i V_i \Lambda(\rho(t_1))V_i^\dagger,
  \end{equation}
   and it is easy to check that
\bea \label{eq:majorization}\aligned
\sum_i V_iV_i^\dagger&=U_2^\dagger (\sum_i K_iK_i^\dagger)U_2=I,\\
\sum_i V_i^\dagger V_i&=U_1^\dagger (\sum_i K_iK_i^\dagger)U_1=I.\\
\endaligned
\eea

It is a linear map from the eigenvalues of $\rho(t_1)$ to eigenvalues of $\rho(t_2)$, so we can find a matrix $D$, such that
\begin{equation}
\lambda(\rho(t_2))=D\lambda(\rho(t_1)),
\end{equation}
where $\lambda(\rho)$ is a vector in $\Re^N$ with eigenvalues of $\rho$ as its entries, which is arranged in the same order as the diagonal entries of $\Lambda(\rho)$. The matrix $D$ can be computed from Eq. (\ref{eq:Kraus-eigen}): $$D_{\alpha\beta}=\sum_i |(V_i)_{\alpha\beta}|^2,$$ where $D_{\alpha\beta}$ and $(V_i)_{\alpha\beta}$ are the $\alpha\beta$ entry of $D$ and $V_i$ respectively. It is straightforward to show that, by using the two conditions in Eq. (\ref{eq:majorization}),
$$\sum_\alpha D_{\alpha \beta}=1,$$
$$\sum_\beta D_{\alpha \beta}=1,$$
i.e., $D$ is a doubly stochastic matrix. From Proposition ~\ref{prop:double}, we get $$\lambda(\rho(t_2))\prec \lambda(\rho(t_1)),$$
so $$\rho(t_2)\prec \rho(t_1), \forall t_2>t_1,$$
i.e. it is majorization monotone.

From Proposition \ref{prop:entropy} and \ref{prop:trace}, it is easy to see that majorization monotone implies entropy monotone and trace norm monotone, and they share the same necessary and sufficient condition: $L(I)=0$.

\section{Conclusion}
Understanding open quantum systems is an important problem for
a wide variety of physics, chemistry, and engineering applications.
This paper analyzed the dynamics of open quantum
systems and gives necessary and sufficient condition on majorization monotone dynamics, which are those dynamics always mixing the states. This suggests that for this class of dynamics, majorization defines an evolution arrow, which begs for the connection to the entropy arrow in the second law of thermal dynamics. I hope further investigation will reveal more on this connection.
%
%


\begin{thebibliography}{10}

\bibitem{Tannor85}
D.~J. Tannor and S.~A. Rice,
\newblock J. Chem. Phys. {\bf 83}, 5013 (1985).

\bibitem{Tannor86}
D.~J. Tannor, R.~Kosloff, and S.~A. Rice,
\newblock J. Chem. Phys. {\bf 85}, 5805 (1986).

\bibitem{Tannor88}
D.~J. Tannor and S.~A. Rice,
\newblock Adv. Chem. Phys. {\bf 70}, 441 (1988).
\bibitem{Kosloff89}
R.~Kosloff, S.~A. Rice, P.~Gaspard, S.~Tersigni, and D.~J. Tannor,
\newblock Chem. Phys. {\bf 139}, 201 (1989).

\bibitem{Rice2000}
S.~A. Rice and M.~Zhao,
\newblock {\em Optical Control of Molecular Dynamics},
\newblock Wiley, New York, 2000.

\bibitem{Shapiro03}
M.~Shapiro and P.~Brumer,
\newblock {\em Principles of the Quantum Control of Molecular Processes},
\newblock Wiley, New York, 2003.

\bibitem{Brixner03}
T.~Brixner and G.~Gerber,
\newblock Chem. Phys. Chem. {\bf 4}, 418 (2003).

\bibitem{Mitric02}
R.~Mitri\'{c}, M.~Hartmann, J.~Pittner, and
V.~Bona\v{c}i\'{c}-Kouteck\'{y},
\newblock J. Phys. Chem. A {\bf 106}, 10477 (2002).

\bibitem{Peirce88}
A.~P. Peirce, M.~A. Dahleh, and H.~Rabitz,
\newblock Phys. Rev. A {\bf 37}, 4950 (1988).

\bibitem{Shi91}
S.~Shi and H.~Rabitz,
\newblock Comp. Phys. Commun. {\bf 63}, 71 (1991).
\bibitem{Tannor04}
S. E. Sklarz, D.~J. Tannor and N. Khaneja,
\newblock Phy. Rev. A {\bf 69}, 053408 (2004).

\bibitem{Jakubetz90}
W.~Jakubetz and J.~Manz,
\newblock Chem. Phys. Lett. {\bf 165}, 100 (1990).

\bibitem{Shen94}
H.~Shen, J.~P. Dussault, and A.~D. Bandrauk,
\newblock Chem. Phys. Lett. {\bf 221}, 498 (1994).

\bibitem{Yan93}
Y.~J. Yan, R.~E. Gillilan, R.~M. Whitnell, K.~R. Wilson, and
S.~Mukamel,
\newblock J. Chem. Phys. {\bf 97}, 2320 (1993).

\bibitem{Khaneja01}
N.~Khaneja, R.~Brockett, and S.~J.~Glaser,
\newblock Phys. Rev. A {\bf 63}, 032308 (2001).
\bibitem{Yuan00}
H.Yuan, N.Khaneja,
\newblock System and Control Letters, Vol 55/6 pp501-507 (2006).
\bibitem{Yuan01}
H.Yuan, N.Khaneja,
\newblock Phys. Rev. A {\bf 72}, 040301 (2005).
\bibitem{Yuan02}
H.Yuan, S. Glaser, N.Khaneja,
\newblock Phys. Rev. A {\bf 76}, 012316 (2007).
\bibitem{Yuan03}
H.Yuan, R. Zeier, N.Khaneja,
\newblock Phys. Rev. A {\bf 77}, 032340 (2008).

\bibitem{Hornung}
T.~Hornung, S.~Gordienko, R.~de~Vivie-Riedle, and B.~Verhaar,
\newblock Phys. Rev. A {\bf 66}, 043607 (2002).

\bibitem{Sklarz02.1}
S.~E. Sklarz and D.~J. Tannor,
\newblock Phys. Rev. A {\bf 66}, 053619 (2002).

\bibitem{sklarz02.2}
S.~E. Sklarz, I.~Friedler, D.~J. Tannor, Y.~B.~Band, and
C.~J.~Williams,
\newblock Phys. Rev. A {\bf 66}, 053620 (2002).

\bibitem{Rangan01}
C.~Rangan and P.~H. Bucksbaum,
\newblock Phys. Rev. A {\bf 64}, 033417 (2001).

\bibitem{Tesch01}
C.~M. Tesch, L.~Kurtz, and R.~de~Vivie-Riedle,
\newblock Chem. Phys. Lett. {\bf 343}, 633 (2001).

\bibitem{Palao02}
J.~P. Palao and R.~Kosloff,
\newblock Phys. Rev. Lett. {\bf 89}, 188301 (2002).
\bibitem{Yuan04}
H.Yuan, R. Zeier, N.Khaneja, S. Lloyd
\newblock Phys. Rev. A {\bf 79} 042309 (2009).
\bibitem{Yuan05}
N. Khaneja, et al,
\newblock Phys. Rev. A {\bf 75} 012322 (2007).

\bibitem{Leibscher03}
M.~Leibscher, I.~S.~Averbukh, and H.~Rabitz,
\newblock Phys. Rev. Lett. {\bf 90}, 213001 (2003).

\bibitem{Rabitz}
H.~Rabitz, R.~de~Vivie-Riedle, M.~Motzkus, and K.~Kompa,
\newblock Science {\bf 288}, 824 (2000).

\bibitem{Gordon97}
R.~J. Gordon and S.~A. Rice,
\newblock Annu. Rev. Phys. Chem. {\bf 48}, 601 (1997).

\bibitem{Bartana93}
A.~Bartana, R.~Kosloff, and D.~J. Tannor,
\newblock J. Chem. Phys. {\bf 99}, 196 (1993).

\bibitem{Bartana}
A.~Bartana, R.~Kosloff, and D.~J. Tannor,
\newblock J. Chem. Phys. {\bf 106}, 1435 (1997).

\bibitem{TannorRot}
D.~J. Tannor and A.~Bartana,
\newblock J. Phys. Chem. A. {\bf 103}, 10359 (1999).

\bibitem{Cao97}
J.~S. Cao, M.~Messina, and K.~R. Wilson,
\newblock J. Chem. Phys. {\bf 106}, 5239 (1997).

\bibitem{Gross98}
P.~Gross and S.~D. Schwartz,
\newblock J. Chem. Phys. {\bf 109}, 4843 (1998).

\bibitem{Ohtsuki03}
Y.~Ohtsuki, K.~Nakagami, W.~Zhu, and H.~Rabitz,
\newblock Chem. Phys. {\bf 287}, 197 (2003).

\bibitem{Khaneja03}
N.~Khaneja, T.~Reiss, B.~Luy, and S.~J.~Glaser,
\newblock Journal of Magnetic Resonance {\bf 162}, 311 (2003).

\bibitem{Lindblad}
G.~Lindblad,
\newblock Commun. Math. Phys. {\bf 33}, 305 (1973).

\bibitem{GKW}V. Gorini, A. Kossakowski, and E. C. G. Sudarshan,
\newblock J. Math. Phys {\bf 17}, 821 (1976).

\bibitem{Alicki86}
R.~Alicki and K.~Lendi,
\newblock {\em Quantum Dynamical Semigroups and Applications},
\newblock Springer, New York, 1986.

%
%
%

\bibitem{bacon}
Dave Bacon, Andrew M. Childs, Isaac L. Chuang, Julia Kempe1,5,6,
Debbie W. Leung, and Xinlan Zhou,
\newblock Phys. Rev. A. {\bf 64}, 062302 (2001)
\bibitem{Bhatia97}
R.~Bhatia,
\newblock {\em Matrix analysis},
\newblock Springer, New York, 1997.
\bibitem{Kraus83}
Kraus K,
\newblock{\em States, Effects and Operations: Fundamental Notions of Quantum Theory},
\newblock  Springer, Berlin, 1983.
\bibitem{Uhlmann}
A. Uhlmann,
\newblock{\em Math.-Naturwiss. Reihe}
\newblock 20, 633 (1971).

\end{thebibliography}
\end{document}